# Extremely long transverse optical needle focus for reflective metalens enabled by monolayer MoS$_2$

Zhonglin Li[1,2,†], Kangyu Gao[1,†], Yingying Wang[1,*], Ruitong Bie[3], Dongliang Yang[3], Tianze Yu[3], Renxi Gao[1], Wenjun Liu[1], Bo Zhong[4], and Linfeng Sun[3,*]

[1]Department of Optoelectronic Science, Harbin Institute of Technology at Weihai, Weihai 264209, China

[2]Department of physics, Harbin Institute of Technology, Harbin 150001, China

[3]Centre for Quantum Physics, Key Laboratory of Advanced Optoelectronic Quantum Architecture and Measurement (MOE), School of Physics, Beijing Institute of Technology, Beijing 100081, China

[4]School of Materials Science and Engineering, Harbin Institute of Technology at Weihai, Weihai 264209, China

Author to whom correspondences should be addressed:

yywang@hitwh.edu.cn; sunlinfeng@bit.edu.cn

[†]These authors contributed equally to this work.

**Abstract:**

Line-scan mode facilitates fast-speed and high-throughput imaging with developing a suitable optical transverse needle focus. Metasurface with periodic structures such as diffractive rings, ellipses, and gratings could enable discrete focus evolving into line focus under momentum conservation, but still face the challenge of extremely low light power utilization brought by inevitably multiple high-order diffractions. In addition, the designed focus requires the selection of particular optical functional materials. High dielectric constants in atomic transition metal dichalcogenides make significant phase modulation by bringing phase singularity at zero-reflection possible. However, no light power is available for use at zero-reflection and a balance between phase and amplitude modulation is needed. In this work, above issues are simultaneously solved by designing a monolayer $MoS_2$ based Fresnel strip structure. An optical needle primary focus with a transverse length of 40 μm (~80 λ) is obtained, which is the longest value recorded so far, together with a sub-diffraction-limited lateral spot and a broad working wavelength range. This specially developed structure not only concentrates light power in primary diffraction by breaking restriction of momentum conservation, but also guarantees a consistent phase across different strips. The novel optical manipulation way provided here together with the longer focus length for flat optics will show promising applications in biology, oncology, nanofabrication, energy harvesting, and optical information processing.

## 1. Introduction

Fast-speed and high-throughput imaging has been extensively used in the fields of biology, oncology, nanofabrication, energy harvesting, and optical information processing. The traditional point-scan mode with insufficient image throughput limits the speed of volumetric imaging.[1-3] Instead, with developing a suitable transverse needle focus, the line-scan mode is able to achieve high imaging throughput by capturing a series of points simultaneously,[4,5] which has been applied in multiple spectroscopies including light-sheet microscopy,[6-10] selective plane-illumination microscopy,[11] and light-sheet tomography,[5,12] realizing the fast acquisition rate up to 100 fps,[13] and capturing at least one or two-order more imaging points compared to a conventional optical system.[14]

In order to access controllable optical field for image scanning, metasurface-based diffractive optical elements with periodic structures are developed which make the realization of any artificial optical filed by tuning light-matter interactions in momentum space under restriction of momentum conservation possible. After patterning functional surface into one-dimensional (1D) or two-dimensional (2D) periodic structures, reciprocal lattice vectors $\vec{G}$ ($\vec{G} = 0, \pm\frac{2\pi}{a}, \pm\frac{4\pi}{a}$, and so on) are generated based on primitive lattice vector **a**. The wave vector of scattering light is determined by the operation of incident light wave vector and reciprocal lattice vector, leading to multiple guided resonance mode for reflective and transmitted light.[15-18] The conservation of momentum is also helpful to re-examine the extremely longitudinal needle focus (large depth of focus) realized by multiple concentric rings from a different view point. For periodic structures with no more than one primitive lattice vectors, a lot of reciprocal lattice vectors are introduced. This makes vector operation of $\vec{G}$ could be variable in a large range along light propagation direction, and the conservation of momentum could be easily satisfied which produces a longitudinal needle focus.[19-21]

Besides needle focus along longitudinal direction, transverse needle focus also could be realized by designing concentric belt structure which generates discrete focal points around the primary spot.[19] Furthermore, by converting the circular functional surface into elliptical belts with sector-shape cutting, a continuous transverse needle focus with a length of 4 μm (7 λ) at 633 nm rather than discrete spots is obtained.[22] Metasurface-based diffractive lens with features of miniaturization, light weighting, and ease of integration, which overcome the limitation of bulky and low integration for traditional cylindrical lens and objective lens in creating transverse optical needle are highly desired.[23] Besides, there are additional challenges for metalens to be overcome includes of breaking the 10 λ needle length, having sub-diffraction-limited lateral spot, a broad working wavelength range, and an easy-to-fabricated structure.

Diffractive grating is a straightforward method in creating transverse optical needle with an extremely long length. However, multiple high-order diffractions besides the primary focus are inevitably for diffractive grating, and the primary focus only receives a small portion of incident light power, which poses a significant obstacle to light power utilization. This is the first issue that should be solved. In addition, the designed focus requires the selection of particularly optical functional materials. Atomic 2D transition metal dichalcogenides (TMDs) present an unprecedented platform for amplitude and phase modulation in flat optics due to their rich dielectric properties and atomic flat surface. To break the limited phase accumulation in atomic TMDs, phase singularity at zero-reflection is introduced to bring a pronounced phase modulation. A large Heaviside-type phase difference of ~π is generated between TMDs and supported substrate in the vicinity of phase singularity.[24-28] However, the large phase modulation is inevitably accompanied by zero-reflection which results in no light power is available for use. Therefore, it is necessary to build a balance between phase modulation and amplitude modulation when using atomic TMDs, and this is the second issue that should be solved.

In this work, after pattering the functional surface polymethyl methacrylate (PMMA)/monolayer $MoS_2$ into 17 Fresnel strips, a transverse needle primary focus with an extremely long length of 40 μm (~80 λ) is reported for the first time. The key to achieve a milestone performance is this specially developed structure although it contains only a few number of Fresnel strips. Those Fresnel strips not only concentrate optical power in primary diffraction by breaking the restriction of momentum conservation, but also ensure a consistent phase across different strips and generate a constructive interference at the focal point. After introducing phase singularity at 616.5 nm, phase difference of 0.2 π-1.0 π and reflectance of 0.1-0.42 are realized for monolayer $MoS_2$ based reflective metalens, which facilitates sub-diffraction-limited lateral spots (0.48 λ/NA-0.62 λ/NA, NA is the numerical aperture) and a wide operating wavelength range (400 nm-800 nm). The designed structure here will promote the development of line-scan spectroscopy, which opens up new avenues for various multidisciplinary applications including biology such as living tissue and whole-brain imaging,[5] oncology such as rapid pathology examination for tumor resection,[4, 29-32] nanofabrication such as line patterns by laser irradiation, energy harvesting, and optical information processing.

## 2. Results and discussions

### 2.1. Phase and amplitude modulation based on multilayer structure

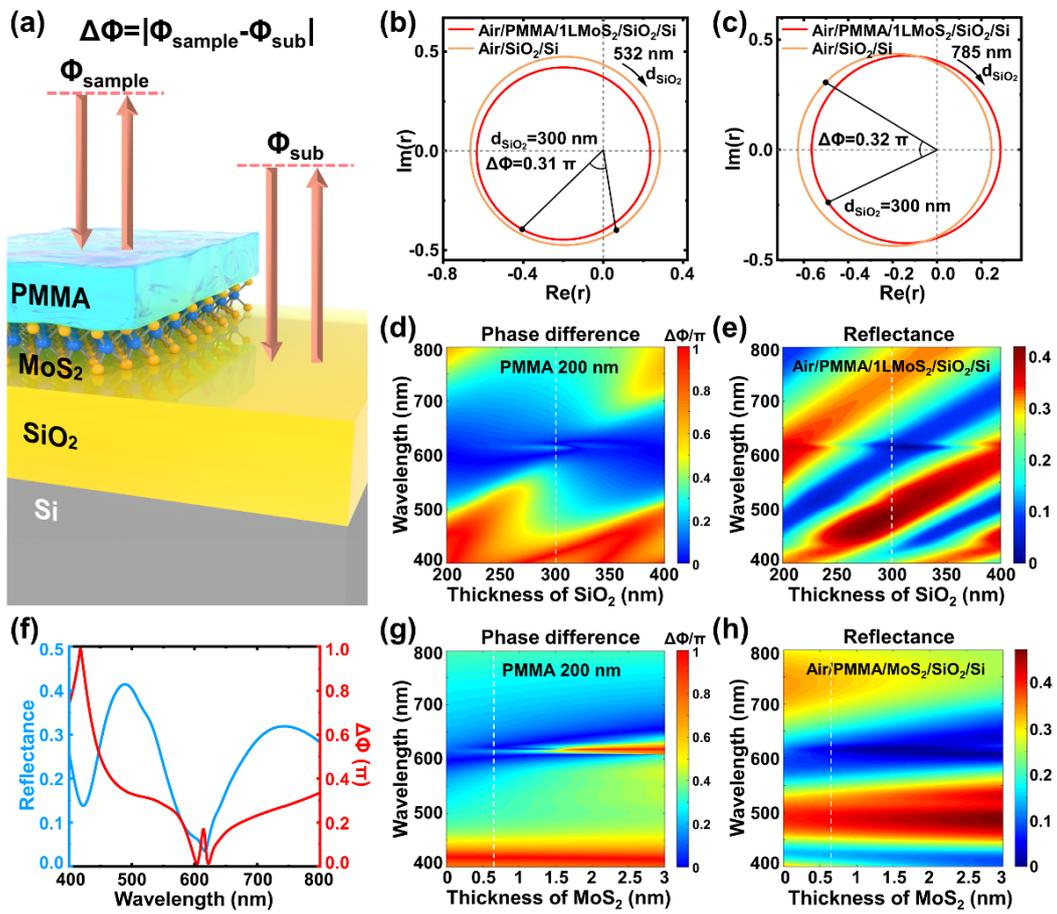

**Fig. 1 PMMA/monolayer MoS$_2$/SiO$_2$/Si multilayer structure is designed to realize efficient phase and amplitude modulation. a** Schematic image of phase difference between functional surface (PMMA/MoS$_2$) and SiO$_2$/Si substrate. **b,c** Phase diagrams of the complex reflections of 200 nm PMMA/monolayer MoS$_2$/SiO$_2$/Si and substrate (SiO$_2$/Si) when SiO$_2$ thickness is swept from 0 nm to 400 nm at wavelengths of 532 nm and 785 nm, individually. **d** The absolute value of phase difference between 200 nm

PMMA/monolayer MoS$_2$ and SiO$_2$/Si substrate as functions of SiO$_2$ thickness and wavelength. **e** The reflectance of 200 nm PMMA/monolayer MoS$_2$ supported on a SiO$_2$/Si substrate as functions of SiO$_2$ thickness and wavelength. 300 nm SiO$_2$ is highlighted by white dashed line. **f** The phase difference between 200 nm PMMA/monolayer MoS$_2$ and SiO$_2$(300 nm)/Si substrate, and the reflectance from 200 nm PMMA/monolayer MoS$_2$/SiO$_2$(300 nm)/Si as a function of wavelength. **g** The absolute value of phase difference between 200 nm PMMA/MoS$_2$ and SiO$_2$(300 nm)/Si substrate as functions of MoS$_2$ thickness and wavelength. **h** The reflectance of 200 nm PMMA/MoS$_2$ supported on a SiO$_2$(300 nm)/Si substrate as functions of MoS$_2$ thickness and wavelength. Monolayer MoS$_2$(0.65 nm) is highlighted by white dashed line.

For atomic materials supported by the SiO$_2$/Si substrate, the high absorptive Si fully absorbs any light transmitted into it, and this reflective system can be treated as one-port resonance system. A Heaviside-type phase jump of π across phase singularity can be realized by bring zero-reflection in one-port resonance system.[24] However, no light power is available for use at zero-reflection, and a balance between phase and amplitude modulation is needed. This is further addressed by introducing a PMMA layer with an optimized thickness, which provides efficient phase and amplitude modulation, improves the mechanical robustness of the system, and protects MoS$_2$ from oxidation and degradation. **Fig. 1a** shows the schematic image of phase difference between PMMA/MoS$_2$ functional surface and supported SiO$_2$/Si substrate. $\Phi_{sample}$, $\Phi_{sub}$, and $\Delta\Phi$ represent the reflective light phase from functional surface, substrate, and their absolute value of difference, respectively. The optical properties including complex reflection coefficient and reflectance of multilayer Fabry-Perot-type interference cavity are calculated from transfer matrix method. **Fig. 1b,c** give the phase diagrams of the complex reflections from 200 nm PMMA/monolayer MoS$_2$(0.65 nm)/SiO$_2$/Si(semi-infinite thickness) and SiO$_2$/Si substrate, when SiO$_2$ thickness is swept from 0 nm to 400 nm at wavelengths of 532 nm and 785 nm, individually. It is found the complex reflection coefficients of functional surface and supported substrate are both on the over-coupling states.[24] There are noticeable phase differences of 0.31 $\pi$ and 0.32 $\pi$ at wavelengths of 532 nm and 785 nm, when SiO$_2$ has a thickness of 300 nm, together with reflectance of 0.32 at 532 nm, and 0.30 at 785 nm from functional surface. The reflectance from the supported substrate and optical contrast are further included in Supplementary Note 1.

**Fig. 1d** gives the absolute value of phase difference between 200 nm PMMA/monolayer MoS$_2$ and SiO$_2$/Si substrate as functions of SiO$_2$ thickness and wavelength. **Fig. 1e** shows the reflectance of 200 nm PMMA/monolayer MoS$_2$ supported on a SiO$_2$/Si substrate as functions of SiO$_2$ thickness and wavelength. **Figs. 1d** and **1e** highlight 300 nm SiO$_2$ by white dashed line. The wavelength-dependent phase difference and reflectance from functional surface (200 nm PMMA/monolayer MoS$_2$ supported on a SiO$_2$(300 nm)/Si substrate) are further depicted in **Fig. 1f**. Phase difference of 0.2 $\pi$-1.0 $\pi$ and reflectance of 0.1-0.42 in the 400 nm to 582.3 nm and 671.5 nm to 800 nm ranges are observed when zero-reflection at 616.5 nm is introduced. Besides, the results of 100 nm PMMA have been given in the Supplementary Note 2. It is found although 100 nm PMMA provides a larger phase difference, however, the extremely low reflectance in a large range is unbenefited for broadening the operating wavelength range. As an optimized parameter, 200 nm PMMA is used in order to make a balance between phase modulation and amplitude modulation. In addition, **Fig. 1g** gives the absolute value of phase difference between 200 nm PMMA/MoS$_2$ and SiO$_2$(300 nm)/Si substrate as functions of MoS$_2$ thickness and wavelength. **Fig. 1h** shows the reflectance of 200 nm PMMA/MoS$_2$ supported on a SiO$_2$(300 nm)/Si substrate as functions of MoS$_2$ thickness and wavelength. It is found this multi-layer structure is able to provide efficient phase difference and reflectance for MoS$_2$ with different number of layers.

## 2.2. The design concept for monolayer MoS$_2$ based reflective metalens

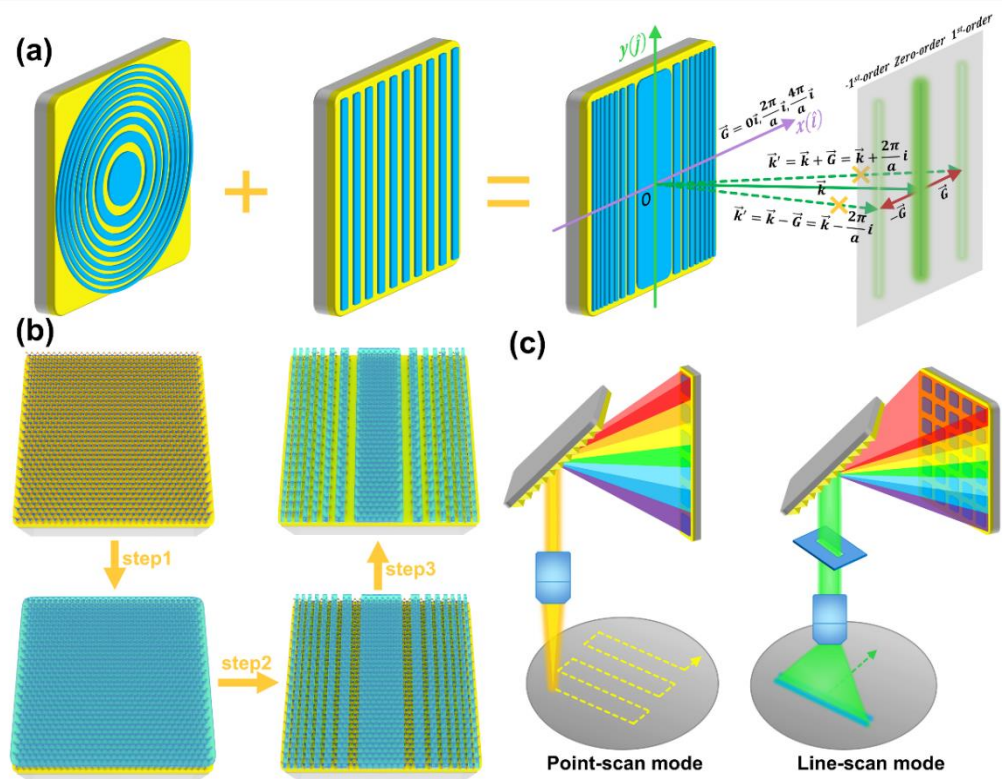

**Fig.2 The design concept for monolayer MoS$_2$ based reflective metalens, which is able to provide an extraordinary long primary transverse needle focus and a high light power utilization. a** The structure is designed based on a combination of a Fresnel zone plate lens and a periodic grating. **b** Schematic image of fabrication process for monolayer MoS$_2$ based reflective metalens. **c** The schematic diagram of the point-scanning mode and the line-scanning mode.

**Fig. 2a** gives the designing concept for monolayer MoS$_2$ based reflective metalens. For a diffractive grating with primary lattice constant $a\vec{i}$, a series of reciprocal lattice vectors is produced as $\vec{G} = 0\vec{i}, \pm\frac{2\pi}{a}\vec{i}, \pm\frac{4\pi}{a}\vec{i}$, and so on. The wave vector of diffractive light including primary order (zero-order), ±1$^{st}$ order, and other high diffractive-order is determined by momentum conservation following $\vec{k}' = \vec{k} + \vec{G}$. Here, $\vec{k}'$ and $\vec{k}$ are wave vectors of diffractive light and incident light, individually. Therefore, the incident light energy is then distributed into a series of diffractions as demonstrated in Supplementary Note 3, which depicts the simulation results of diffractive grating with uniform-width strips at different wavelength. The primary focus only receive a small portion of incident light which severely degrades light power utilization. Differently, when the structure is specifically designed as a combination of a Fresnel zone plate lens and a diffractive grating, the exceptionally long transverse needle focus together with a high light power utilization are feasible. The non-periodic Fresnel-strip structure results in the disappearance of reciprocal lattice vector, which leads to the suppression simultaneously created high-order diffractions at the focal plane. The line focus by concentrating light power on the primary focus with homogeneity and condensation is maintained, as schematically shown in **Fig. 2a**. In addition, the widths of those Fresnel strips are designed based on Fresnel's equation (Supplementary Note 4) which guarantee a constant phase difference across different strips, leading to constructive interference at the focal spot and leading to a high light power utilization.

**Fig. 2b** gives the schematic image of fabrication process for monolayer $MoS_2$ based reflective metalens. Monolayer $MoS_2$ is grown on $SiO_2$(300 nm)/Si substrate by chemical vapor deposition (CVD) method. The 200 nm PMMA photoresist is subsequently spin coated on $MoS_2$. PMMA/monolayer $MoS_2$ film are further patterned into 17 strips with a length of 40 μm using e-beam lithography (EBL, method). A focal length of 45.5 μm is designed for the reflective metalens at 532 nm and the width of every strip is calculated based on the Fresnel's equation considering incident wavelength $\lambda$ and focal length (Supplementary Note 4). **Fig. 2c** shows the schematic diagram of the point-scanning mode and the line-scanning mode, and the transverse optical needle based line-scanning mode significantly boosts the imaging throughput by capturing numerous points simultaneously.

**2.3. Optical performance of the fabricated reflective metalens at 532 nm**

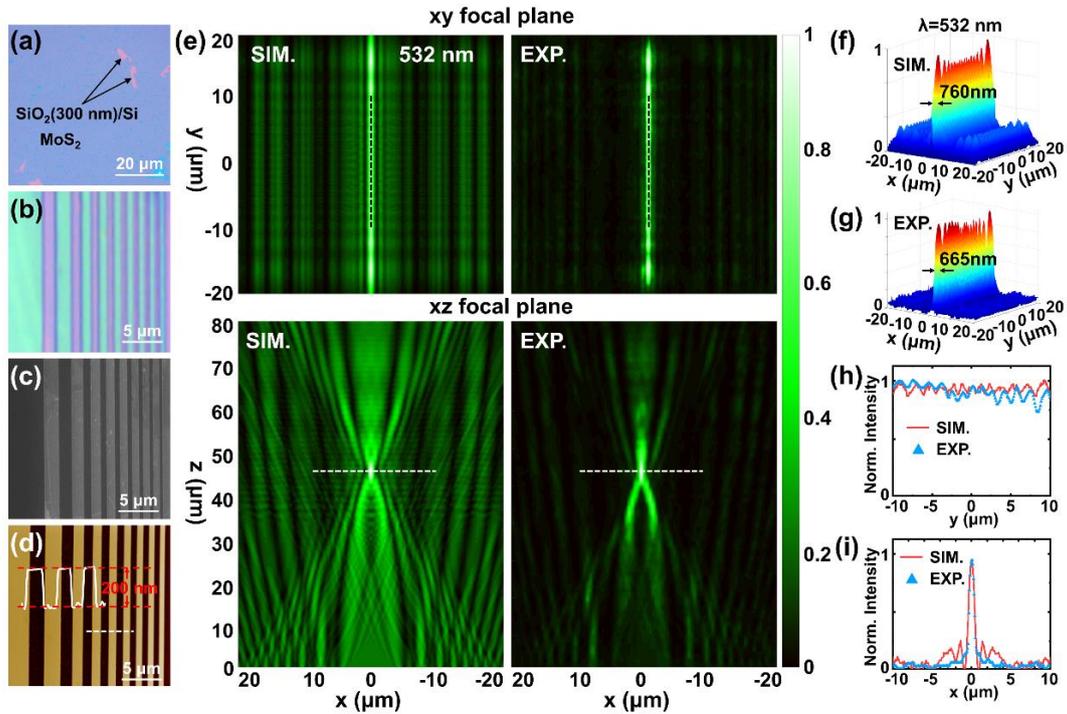

**Fig.3 The focusing properties of the fabricated reflective metalens at 532 nm both**

**theoretically and experimentally. a** Optical image of CVD monolayer MoS$_2$ grown on SiO$_2$(300 nm)/Si substrate. **b-d** Optical image, SEM image, and AFM image of the fabricated reflective metalens. **e** The normalized intensity distributions at 532 nm in the xy focal plane and xz focal plane both theoretically and experimentally. **f,g** The normalized intensity distributions on the xy focal plane both theoretically and experimentally. **h** The normalized intensity along the black dashed lines in **Fig. 3e** both theoretically and experimentally. **i** The normalized intensity along the white dashed lines in **Fig. 3e** both theoretically and experimentally.

**Fig. 3a** displays the optical image of monolayer MoS$_2$ grown on SiO$_2$(300 nm)/Si substrate. Clear optical contrasts are observed for monolayer MoS$_2$ and substrate. **Figs. 3b-d** show the optical image, scanning electronic microscopy (SEM) image, and atomic force microscopy (AFM) image of the fabricated reflective metalens. The functional surface 200 nm PMMA/monolayer MoS$_2$ is patterned into 17 long strips, each with 40 μm in length and varying in width from 0.59 μm to 9.66 μm. The structure parameters for those strips have been given in the Supplementary Note 4. **Fig. 3d** illustrates a ~200 nm height difference between functional surface and supported substrate.

The simulation is then performed by using finite-difference time-domain (FDTD) method based on 17 strips with designed parameter with a length of 40 μm (Method). The functional surface is located on the xoy plane, with light normally incidents and reflects from functional surface, further propagates along z direction. The experimental performances of the fabricated metalens are measured by MetronLens, which show good agreements with theoretical simulations (Method). It is found the fabricated reflective metalens generates a line focus at z=46 μm which is consistent with our designed focal length of 45.5 μm at 532 nm. **Fig. 3e** gives the normalized intensity distributions at 532 nm in the xy focal plane and in the xz focal plane (y=0), both experimentally and theoretically. It is found the designed reflective lens generates a well-defined primary transverse optical needle focus with 40 μm (~80 λ) length in the xy focal plane, together with ignorable created side-lobes (<10% of the intensity of the central primary focus), demonstrating its primary-order focus properties. This specially designed structure indeed promotes the suppression of those simultaneously created side-lobes. The intensity distributions along the needle region at the xy focal plane both theoretically and experimentally are further given in **Fig. 3f,g**, while the line intensity profiles along the transverse optical needle at x=0 are presented in **Fig. 3h**. The slight variations in intensity experimentally and theoretically may be originated from the fabrication imperfections and the limited number of grids used in the simulation. It should be noted that the length of the transverse needle is limited by the size of the grown $MoS_2$, and a wafer-scale $MoS_2$ will generate a transverse needle focus with a length of few inches.

The sub-diffraction-limited lateral focus is further investigated both experimentally and theoretically and the results are shown in the lower panels in **Fig.3e**, together with the line intensity profiles perpendicular to the transverse optical needle are given in **Fig. 3i**. The experimental lateral focal spot of 665 nm throughout the entire optical needle region is observed, which shows consistent with the theoretical value of 760 nm. The obtained lateral focal spot of 665 nm (~0.50 $\lambda$/NA) at 532 nm is smaller than that of Rayleigh Criterion (0.61 $\lambda$/NA), demonstrating the sub-diffraction-limited focal property within the needle region for this fabricated metalens.

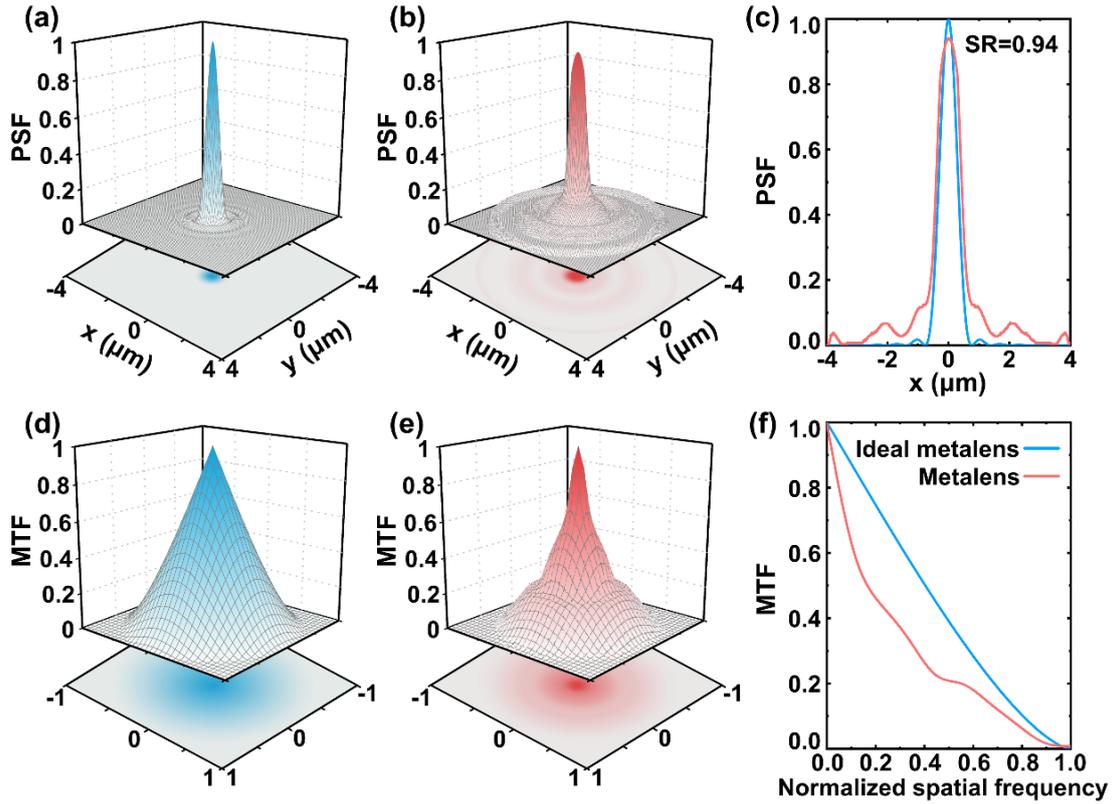

**Fig.4 The PSF and MTF results of the fabricated reflective metalens at 532 nm. a-c** PSF results of the ideal lens (blue color) and the fabricated metalens (red color), and SR value of 0.94 is obtained. **d,e** MTF results of the ideal lens (blue color) and fabricated metalens (red color). **f** MTF results of the ideal lens and fabricated metalens as a function of normalized spatial frequency.

The phase distribution for the fabricated metalens is further measured using MetronLens (Method), which allows for the acquisition of optical performance including pupil function, points spread function (PSF) and modulation transfer function (MTF).[29] The definitions of those parameters are given in Supplementary Note 5. The experimental PSF and MTF results of the fabricated metalens at 532 nm are accessed using an experimental lateral focal spot of 665 nm. The PSF value can be used to evaluate how an optical system response to a point object, while a point object is blurred into a three-dimensional (3D) image with a certain shape and size after reflection from lens. **Fig. 4a,b** show the PSF distributions on the focal plane for the ideal lens (blue color) and the fabricated metalens (red color) at 532 nm. The PSF values of the ideal lens are calculated based on ideal phase distribution as given in Supplementary Note 5. SR value of 0.94 is obtained experimentally, which is the peak value of the measured PSF value at the exit pupil normalized to the peak value of ideal lens as shown in **Fig. 4c**. Based on the Marechal standard, a lens with a SR>0.8 is considered to be diffraction-limited with tolerable wave aberration. While for ideal lens without aberration, the SR value is 1.

By analyzing MTF in spatial frequency range, the abilities of this metalens in transferring object detail, object contour, and object hierarchy are further evaluated. Normally, MTF values drops to 0 at a cutoff spatial frequency of $2NA/\lambda$, with the largest spatial frequency limited by values of NA and wavelength. **Fig. 4d,e** show the MTF distributions on the focal plane of ideal lens (blue color) and the fabricated metalens (red color) at 532 nm. **Fig. 4f** presents the MTF values both for the fabricated metalens and the ideal lens (calculated from the ideal phase distribution as given in Supplementary Note 5), with normalized spatial frequency. As can be seen, the MTF of the fabricated metalens in the high frequency range is close to that of a perfect lens, demonstrating its super object detail transfer ability.

**2.4. The simulation results of the reflective metalens in a wide operating wavelength range**

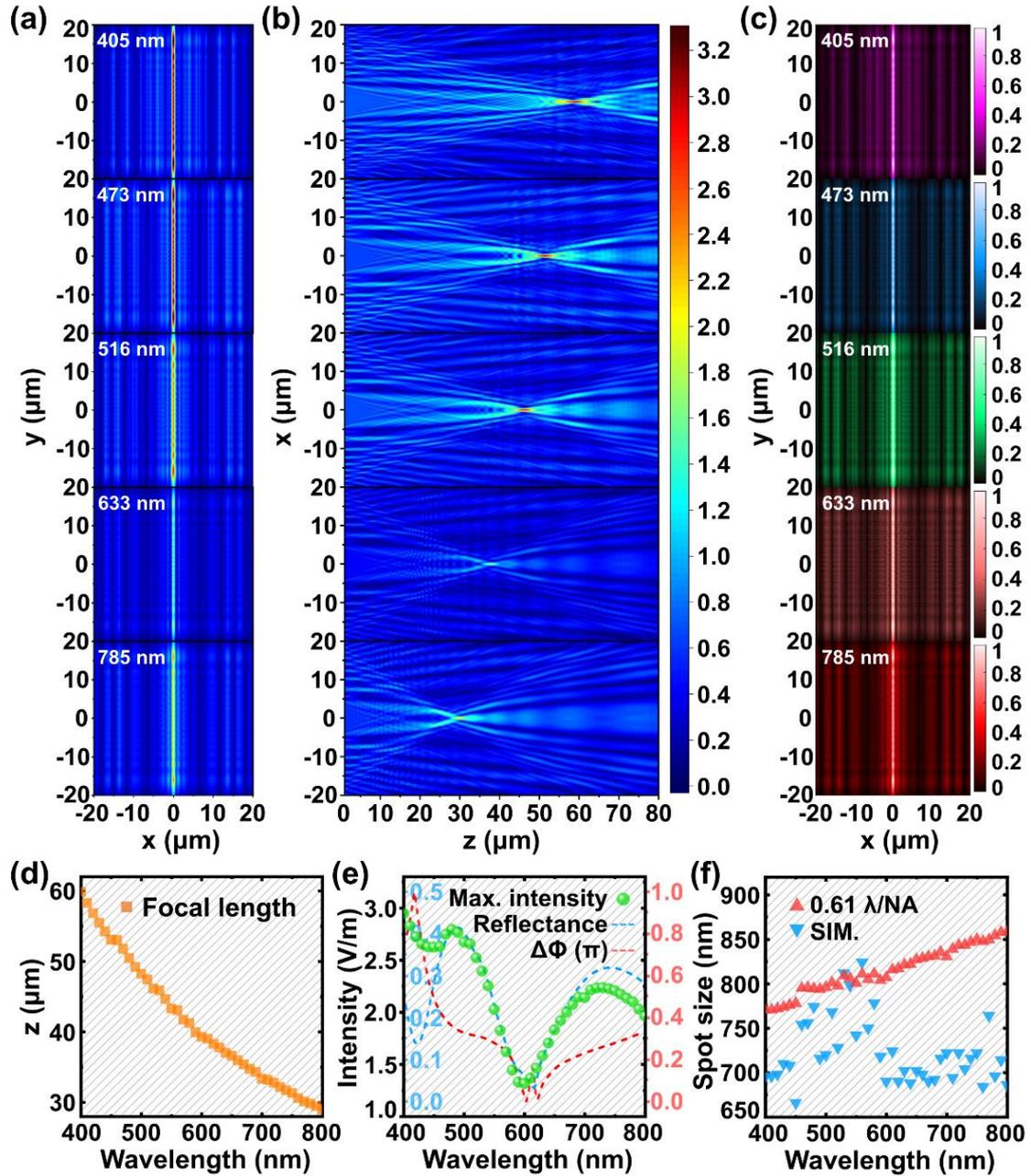

**Fig.5 Simulation results of intensity distributions at the focal plane in a wide operating wavelength range. a,b** Light field intensity distributions in the xy focal plane and xz focal plane for different wavelength. **c** The pseudo-color normalized intensity distribution in the xy focal plane for different wavelength. **d** The focal length of the reflective metalens at different wavelength. **e** The maximum electric field intensity at the focal plane for different wavelength. **f** The lateral spot of the focus at different wavelength and compared with those of Rayleigh Criterion.

Even though the reflective metalens is designed for focusing 532 nm, it is discovered that this specifically designed structure is able to operate in a wide range of wavelengths. X-polarized parallel light with an electric filed intensity of 1 V/m at different wavelength is normally incident on this metalens, and the reflective light field is systematically simulated. The light intensity distributions in the xy focal plane and in the xz focal plane in a wide range of 400 nm to 800 nm are depicted in **Fig. 5a,b**. Extremely long transverse optical needles with lengths of 40 μm are observed for different wavelength with constant intensities along the needle region. **Fig. 5c** shows the pseudo-color normalized intensity distributions in the xy focal plane for different wavelength. This reflective metalens enables the generation of a long transverse primary needle focus with low intensity side-lobes (10%-18% of the primary focus) in a wide wavelength range.

The negative dispersion of this reflective metalens is further illustrated in **Fig. 5d**, and with the increase of wavelength, the focal length is decreased. **Fig. 5e** presents the maximus electric field intensity at the focal plane at different wavelength, along with comparisons of phase difference and reflectance (as given **Fig. 1f**). It can be seen, the maximus electric field intensity resembles the phase difference and reflectance very well. Besides, focusing efficiency at different wavelength is further calculated and given in Supplementary Note 6, which also follows the patterns of phase difference and reflectance very well. Therefore, it demonstrates in order to achieve a high optical performance, an optimized modulation in both phase and amplitude is needed for atomic TMDs. In addition, benefited from high dielectric constants of TMDs over a large wavelength range, broad-band sub-diffraction-limited lateral spots are further observed in **Fig. 5f**. Lateral focal spots ranging from 666 nm to 900 nm (0.48 λ/NA-0.62 λ/NA) are observed at different wavelength, demonstrating its sub-diffraction-limited focus properties within the needle region across a wide wavelength range.

### 3. Conclusion

In summary, a functional surface with 17 Fresnel strips based on monolayer $MoS_2$ is designed and fabricated. Unlike a traditional diffractive grating which scatters light into multiple high-diffraction order, this specifically designed structure breaks the restriction of momentum conservation and concentrates incident light power into the primary focus, which greatly improves light power utilization. It is found the reflective metalens generates an optical needle focus with an exceptionally long transverse length of 40 μm (~80 λ), sub-diffraction-limited lateral spots of 0.48 λ/NA-0.62 λ/NA, and a broad operating wavelength of 400 nm-800 nm. The incredibly long transverse optical needle focus with lateral spots below Rayleigh Criterion, the straightforward simple surface structure, as well as the scalable fabrication process obtained here will promote the development of line-scan mode based imaging spectroscopy with fast-speed and high-throughput, and this will enable tremendous multidisciplinary applications in biology, oncology, nanofabrication, energy harvesting, and optical information processing.

### 4. Method and experimental section

**The fabrication of the reflective metalens**: Monolayer $MoS_2$ was grown by CVD method on $SiO_2$/Si substrate. 950 PMMA photoresist was spin coated on monolayer $MoS_2$ at a speed of 4000 rpm for 1 min. The PMMA coated monolayer $MoS_2$ was then baked on a hot plate at 120 °C for 5 min in order to promote the solidification of the PMMA layer. The PMMA layer was patterned into the designed structure by using an EBL system (eLINE Plus, Raith). The PMMA coated film was further immersed into a developer solution for 2 min (with concentration of MIBK: IPA = 1: 3), and finally immersed into a fixer solution for 40 s (pure IPA solution). The exposed region of the PMMA layer was completely resolved into solution, yielding the designed structure. Subsequently, the structure was further subjected to a 20-second plasma etching treatment (SHL 100μ-RIE, Beijing, Sanhelian) with an etching power of 40 W at an argon gas flow of 10 sccm. This resulted the complete removing of the exposed monolayer $MoS_2$.

**The simulation of the reflective metalens**: The focus properties of the reflective metalens are simulated by the FDTD software using a parallel beam of x-polarized light rays incident vertically on the functional surface. The reflective light from the functional surface was simulated and the intensity distributions in the focal region are obtained.

**The experimental measurements of the reflective metalens**: The experimental performance of the reflective metalens was measured using MetronLens-Phi from Ideaoptics Co., Ltd, Shanghai. For the intensity profile measurements, a linear polarized 532 nm laser was focused by an objective lens with a NA of 0.75 to illuminate the fabricated reflective metalens, and the reflective beam modulated by the reflective lens was collected by the same objective. The diffraction pattern was collected by a high quantum efficiency BSI sCMOS detector. For the phase distribution measurements, an off-axis digital holography technique was adopted, where the reference beam was directed at an angle to the sample beam, which enabled the recording of interference patterns. The interference patterns were sequentially underwent Fourier transformation, filtering, and inverse Fourier transformation processes to reconstruct the phase distribution modulated by the fabricated lens. The resolution of phase measurement was 50 mrad with a spatial resolution of 0.6 μm by using a 40x objective lens, and the maximum measured focal length range was 5 mm.

## Acknowledgements

L. Sun acknowledges the support from the Beijing Natural Science Foundation (Grant No. Z210006), National Natural Science Foundation of China (Grant No. 12104051), and National Key R&D Plan (2022YFA1405600). This work was supported by the Fundamental Research Funds for the Central Universities.

## Author contributions

Z.L. Li and K.Y. Gao contributed equally to this work. Y.Y. Wang and L.F. Sun conceived the idea, supervised the project, and revised the manuscript. Z.L. Li and R.T. Bie fabricated the reflective metalens structure. R.X. Gao, W.J. Liu, and B. Zhong provided discussions in the simulation, and K.Y. Gao performed the simulation. Z.L. Li, R.T. Bie, D.L. Yang, and T.Z. Yu performed the experimental measurements. Z.L. Li and K.Y. Gao performed data analyzation and manuscript preparation. All authors contributed to the discussion.

## Conflict of Interest

The authors declare no competing interests.

## Data Availability Statement

The data within this paper are available from the corresponding author upon reasonable request.